\documentclass[english]{sig-alternate}
\usepackage[T1]{fontenc}
\usepackage[latin9]{inputenc}
\usepackage{float}
\usepackage{graphicx}


\usepackage{babel}
\begin{document}

\title{SocialDTN: A DTN implementation\\
for Digital and Social Inclusion % You need the command \numberofauthors to handle the 'placement
\numberofauthors{1} %  in this sample file, there are a *total*
}

\author{% You can go ahead and credit any number of authors here,
% 1st. author
\alignauthor Waldir Moreira$^{1}$, Ronedo Ferreira$^{2}$, Douglas
Cirqueira$^{2}$,\\
Paulo Mendes$^{1}$ and Eduardo Cerqueira$^{2,3}$\\
\affaddr{$^{1}$SITILabs, University Lusófona, $^{2}$ITEC, Federal
University of Pará, $^{3}$CSD, UCLA}\\
 \affaddr{waldir.junior@ulusofona.pt, ronedo@aitinet.com, douglas.cirqueira@gmail.com,
paulo.mendes@ulusofona.pt, cerqueira@ufpa.br} }
\maketitle
\begin{abstract}
Despite of the importance of access to computers and to the Internet
for the development of people and their inclusion in society, there
are people that still suffer with digital divide and social exclusion.
Delay/Disruption-Tolerant Networking (DTN) can help the digital/social
inclusion of these people as it allows opportunistic and asynchronous
communication, which does not depend upon networking infrastructure.
We introduce SocialDTN, an implementation of the DTN architecture
for Android devices that operates over Bluetooth, taking advantages
of the social daily routines of users. As we want to exploit the social
proximity and interactions existing among users, SocialDTN includes
a social-aware opportunistic routing proposal, \emph{dLife}, instead
of the well-known (but social-oblivious) \emph{PROPHET}. Simulations
show the potential of \emph{dLife} for our needs. Additionally, some
preliminary results from field experimentations are presented.
\end{abstract}
%A category including the fourth, optional field follows...


 % NOT required for Proceedings

\section{Introduction}

The number of computers and Internet users increase everyday as they
play an important role not only in the educational development of
people, but also in the inclusion of people into society. However,
due to being geographic dispersed from the major cities (e.g., Amazon
regions), some people are completed left out of the digital world.

Due to its capability of coping with intermittent connectivity and
long delays, and due to the employment of the store-carry-and-forward
paradigm, DTN can be used to mitigate this digital divide. It can
allow users of isolated regions (e.g., riverside communities of the
Amazon region) to opportunistically and asynchronously access the
Internet and communicate with other users inside and outside such
regions.

One can find different implementations of the DTN architecture \cite{rfc4838}
that can be employed in such scenario, such as DTN2, IBR-DTN and Bytewalla.
However, as social excluded communities do not count with any network
infrastructure, but still present social proximity and interactions
patterns (e.g., due to daily habits and routines), these implementations
do not answer our needs as they do not operate over Bluetooth in order
to capture social proximity, do not employ routing approaches able
to exploit social interation patterns, and depend on some infrastructure
(e.g., AP) to allow user devices to communicate.

Thus, we present SocialDTN, an implementation of the DTN architecture
and Bundle Protocol \cite{rfc5050} for Android devices that takes
advantage of the social proximity and daily routines of users to exchange
bundles, even in the absence of any network infrastructure. In order
to exploit social proximity and interactions between users, SocialDTN
introduces a Bluetooth Convergence Layer aiming to consider only socially
well-connected users (i.e., devices) in the exchange of bundles, in
order to increase the probability of message delivery while avoiding
waste to network and storage resources. Social proximity relates to
how physically close users are, and how much time they spend together.
This means that Bluetooth is the most indicated technology to exploit
such proximity as it supports only close range communication. Additionally,
the exchange of bundles solely between socially well-connected users
is achieved with \emph{dLife} \cite{dlife,draftDlife}, which computes
the social weight among users considering their social interactions
and the importance that users have in different periods of their daily
routines. 

We present some simulation results considering the social-aware \emph{dLife
}\cite{dlife} and the social-oblivious \emph{PROPHET} \cite{prophet}
in a trace scenario aiming to show how useful social proximity is
for SocialDTN. Also some preliminary results of SociaDTN based on
field experiments are presented.

This work is part of a joint project between SITILabs and Federal
University of Pará (UFPA) called DTN-Amazon. This project aims to
develop networking technology to mitigate the effects of digital divide
and social exclusion in the riverside communities close to the UFPA
campus in Belém, Pará, Brazil. 

The remaining of this paper is structured as follows: We present the
SocialDTN implementation in Section 2. Section 3 provides a description
of the simulation and field experimentation results. On Section 4
we conclude our arguments and and present some future steps of the
DTN-Amazon project.

\section{SocialDTN\label{sec:The-OppNetModule}}

\begin{figure}[H]
\begin{centering}
\includegraphics[scale=0.3]{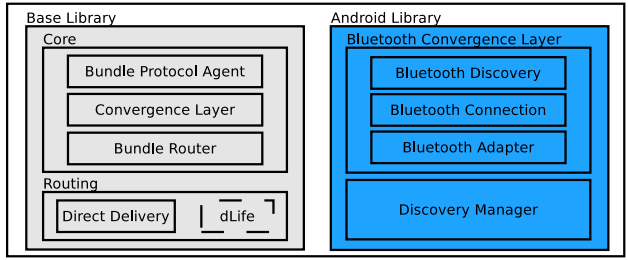}
\par\end{centering}

\protect\caption{\label{fig:Opportunistic-Networking-Module}Overview of SocialDTN}
\end{figure}

Fig. \ref{fig:Opportunistic-Networking-Module} presents the SocialDTN
implementation. It implements the DTN architecture based on RFC\,4838
\cite{rfc4838} and RFC\,5050 \cite{rfc5050} for basic DTN and bundle
exchange operations. As mentioned earlier, current implementatons
of the DTN archictecture (i.e., DTN2, IBR-DTN, Bytewalla) do not comply
with our main requirements: i) exploit social proximity and interactions
between users; and ii) be independent of any network infrastructure.
With SocialDTN social proximity is exploited by using Bluetooth as
it is a close-range communication technology, but none of such implementations
are functional over this wireless technology. What is more, these
implementations depend somewhat of some infrastructure (e.g., an AP)
to aid the communication between endpoints, which is not feasible
(due to its nonexistence) for the targeted communities. 

Thus, the Bluetooth Convergence Layer (BCL) was implemented to allow
SocialDTN to operate over the Bluetooth technology and to allow exploitation
of the aforementioned social proximity. With BCL, nodes can straightforwardly
exchange bundles through the Bluetooth interface without the need
of a structured WiFi network. SocialDTN takes advantages of some native
Bluetooth functionality already implemented in Android, such as the
discovery of neighbors and the possibility of storing information
about them: BCL comprises a Service Discovery Protocol (SDP) for sensing
the medium and the Serial Port Profile (SPP) for data exchange. Moreover,
it uses RFCOMM as transport protocol and runs on top on the Logical
link control and adaptation protocol (L2CAP), which interfaces with
the Host Controller Interface (HCI). 

Since we want the exchange of bundles to happen mostly between socially
well-connected users, we want to make use of these social interactions.
This leads us to the opportunistic routing component, which is responsible
for deciding what is the best way for a bundle to reach its destination.
The routing agent of SocialDTN is implemented based on the specification
of the social-aware opportunistic proposal \emph{dLife} \cite{dlife,draftDlife}.
It is important to note that \emph{PROPHET} \cite{prophet} could
have been used; however, it is a social-oblivious proposal%
\footnote{Some authors may argue that \emph{PROPHET} is a social-aware solution
as it considers the history of encounters. However, we believe that
social-aware solutions are much more elaborated and defined based
on utility functions able to identify/classify individuals or groups
of these, i.e., interests, social ties, levels of social interactions.%
}, and does not explore the social interactions as we intend to use
in SocialDTN. In Section 3, we show why \emph{dLife} answers our needs
better than \emph{PROPHET}.

\section{Results\label{sec:Evaluation-and-Results}}

This section is divided into two parts: first, to show that the social-aware
\emph{dLife} \cite{dlife,draftDlife} is able to perform better than
the social-oblivious \emph{PROPHET} \cite{prophet}; second, to show
our preliminary field experiments.

\subsection{Social-aware vs. Social-oblivious}

We want to exploit social proximity and interactions with SocialDTN.
Thus, we carried out simulations on the Opportunistic Network Environment
(ONE) simulator using the CRAWDAD traces of Cambridge that comprises
contact information of 36 students carrying these devices throughout
their daily routines \cite{cambridge-haggle-imote-content-2006-09-15}.
Our goal is to show that \emph{dLife} can capture this social proximity,
and can perform better than \emph{PROPHET} \cite{prophet} justifying
our choice for the former for routing data in socially excluded communities,
which have similarities to a university small community without Internet
access (e.g. short geographic dispersion, small number of people).

A total of 6000 messages are generated with size ranging from 1 kB
to 100 kB. The source/destination pairs remain the same for the simulations
of both proposals. Since different applications (e.g., email, asynchronous
chat) are expected to run over the proposed module, we set message
Time-To-Live (TTL) in 1, 2, 4 days, 1 week and 3 weeks. The results
are presented with 95\% confidence interval and in terms of average
delivery probability (ratio between the number of delivered messages
and total number of created messages) and cost (number of replicas
per delivered message).

\begin{figure}
\begin{centering}
\includegraphics[scale=0.7]{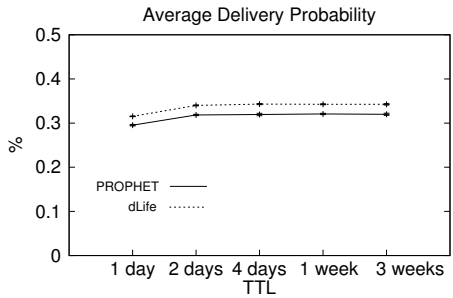}
\par\end{centering}

\protect\caption{\label{fig:Delivery-performance}Delivery performance with human traces}
\end{figure}

\begin{figure}
\begin{centering}
\includegraphics[scale=0.7]{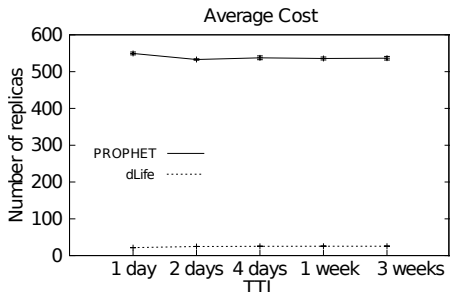}
\par\end{centering}

\protect\caption{\label{fig:Cost-performance}Cost performance with human traces}
\end{figure}

In Figure. \ref{fig:Delivery-performance} and Figure \ref{fig:Cost-performance},
we can observe that \emph{dLife} has better overall performance than
\emph{PROPHET}. Regarding delivery, both proposals are affected due
to the very low number of contacts (average 32 per hour) in the scenario.
Thus, determining social weight/node importance (\emph{dLife}) and
delivery predictability of nodes (\emph{PROPHET}) takes longer to
offer both proposals a stable view of the network in terms of the
utility functions that both use. Yet, \emph{dLife} has a subtle advantage
over \emph{PROPHET}. 

When it comes to cost, clearly we see the advantage of relying on
social awareness. The advantage of \emph{dLife} over \emph{PROPHET}
is of approximately 25 times better: \emph{dLife} creates an average
of 24.56 replicas to perform a successful delivery, while \emph{PROPHET}
creates 538.37 replicas. 

With these results, we can see that \emph{dLife} indeed captures the
social proximity we want to exploit in the context of DTN-Amazon:
the lower number of replicas suggests that only well socially-connected
nodes exchange content. This allows us to have content easily going
from the UFPA campus to the riverside communities through nodes which
are really involved in the the process of digital/social inclusion.

\subsection{Field Experimentations}

SocialDTN first deployment tests consider seven Android devices carried
by students during their daily routine activities in the UFPA campus
for five days. A traffic generator was implemented to run in each
device and to create a load of 6 messages/hour, towards the other
six nodes present in the experiments. Nodes counted with a 10MB storage
and message size varied between 1KB and 1MB as to represent the different
applications and data types (e.g., asynchronous chats, email, video
sharing) that are expected to be used. 

The goal of these tests are to: i) evaluate the BCL implementation
and generate the first DTN-Amazon contact traces; and ii) test if
the implementation of \emph{dLife} complies with its specification
\cite{draftDlife}. Both BCL and \emph{dLife }have shown to be sound:
with the former we are able to detect the social proximity among nodes
and explore their social interactions with the latter. Additionally,
we can collect trace information that can be used in the ONE simulator.

\begin{figure}
\begin{centering}
\includegraphics[scale=0.3]{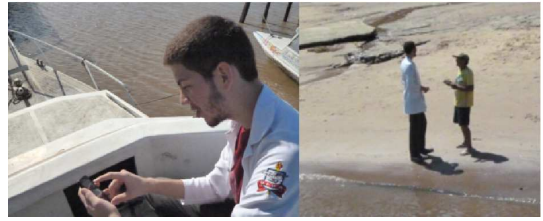}
\par\end{centering}

\protect\caption{\label{fig:agent}SocialDTN usage by Riverside community}
\end{figure}

Another experimentation (cf. Fig. \ref{fig:agent}) involved a health
agent that acts in the riverside community of Combu Island. As SocialDTN
``knows'' this agent has a strong interaction towards this community,
the device he carried received a video%
\footnote{https://www.youtube.com/watch?v=JUDZ8hMnZeI%
} about dengue prevention while he was visiting the Bettina Ferro hospital
in the UFPA campus. On the way to the community, the agent checks
for any information he should be using in his visit. The ability of
using videos does increase the efficiency of the agent actions as
the population in this area is illiterate.

\section{Conclusions and Future Work\label{sec:Conclusions}}

This work presents a new instantiation of the DTN architecture, called
SocialDTN, which currently implements RFC 4838 and RFC 5050, as well
as a new Bluetooth Convergence Layer and a new opportunistic routing
protocol, \emph{dLife}. SocialDTN is being developed in the context
of the DTN-Amazon project, aiming to promote the social/digital inclusion
of the Amazon riverside communities in the vicinity of the UFPA University
campus.

As shown by our experimental work, SocialDTN is able to explore the
social proximity and interactions and to operate with no dependence
upon network infrastructure.

As future work, we plan to: i) have BCL supporting multiple connections;
ii) include other opportunistic routing solutions, such as \emph{Epidemic}
and \emph{PROPHET}; iii) expand the experimentations in the riverside
communities; iv) release the collected social traces to the scientific
community.

\section*{Acknowledgment}

\noindent Thanks are due to the National Council for Scientific and
Technological Development (CNPq), to the Amazon Research Foundation
(FAPESPA), and to FCT for financial support of the User-Centric Routing
project (PTDC/EEA-TEL/103637/2008) and Waldir Moreira's PhD grant
(SFRH/ BD/62761/2009).

\bibliographystyle{ieeetr}
\bibliography{bib-or}

\end{document}